\documentclass[traditabstract]{aa} 
\usepackage{amsmath}

\newcommand{\bea}{\begin{eqnarray}}
\newcommand{\eea}{\end{eqnarray}}
\newcommand{\be}{\begin{equation}}
\newcommand{\ee}{\end{equation}}
\newcommand{\bc}{\begin{center}}
\newcommand{\ec}{\end{center}}
\newcommand{\ben}{\begin{enumerate}}
\newcommand{\een}{\end{enumerate}}
\newcommand{\bd}{\begin{description}}
\newcommand{\ed}{\end{description}}
\newcommand{\bmi}[1]{\begin{minipage}{#1 cm}}
\newcommand{\emi}{\end{minipage}}
\newcommand{\bmif}[1]{\begin{minipage}{#1\textwidth}}

\usepackage{graphicx}

\usepackage{color}
\usepackage{txfonts}
\usepackage{rotating}
\usepackage{multirow}
\usepackage{url}
\usepackage{txfonts}
\usepackage{natbib}
\usepackage{subfigure}
\usepackage{enumerate}

\newcommand{\avg}[1]{\left< #1 \right>} 

\def\OIII{[\ion{O}{iii}]}

\def\Hbeta{\ion{H}{$\beta$}}
\def\Halpha{\ion{H}{$\alpha$}}

\def\SiIV{\ion{Si}{iv}}

\def\OIV{\ion{O}{iv}}
\def\NV{\ion{N}{v}}
\def\SiIV{\ion{Si}{iv}}

\def\Lyalpha{\ion{Ly}{$\alpha$}}
\def\CIII{\ion{C}{III]}}
\def\CIV{\ion{C}{iv}}
\def\PV{\ion{P}{v}}
\def\CIVd{\ion{C}{iv}\,$\lambda$1549}

\def\obj{H1413$+$117}

\begin{document}

   \title{Evidence for two spatially separated UV continuum emitting regions in the Cloverleaf broad absorption line quasar\thanks{Based on observations made with ESO Telescopes at the Paranal Observatory (Chile). ESO program
                ID: 386.B-0337}}
 \author{D. Sluse
 \inst{1} 
 \and  D. Hutsem\'ekers
 \inst{1}
\and T. Anguita
\inst{2,3}
\and  L. Braibant
\inst{1}
\and P. Riaud
\inst{4}
 }

 \institute{Institut d’Astrophysique et de G\'eophysique, Universit\'e de Li\`ege, Quartier Agora - All\'ee du six Ao\^ut, 19c , B-4000 Li\`ege, Belgium
 \email{dsluse@ulg.ac.be}
 \and Departamento de Ciencias Fisicas, Universidad Andres Bello, Fernandez Concha 700, Las Condes, Santiago, Chile
 \and Millennium Institute of Astrophysics, Chile
 \and Rue des Bergers, 60, F-75015 Paris, France}

\date{Received ; accepted}

  \abstract {Testing the standard Shakura-Sunyaev model of accretion is a challenging task because the central region of quasars where accretion takes place is unresolved with telescopes. The analysis of microlensing in gravitationally lensed quasars is one of the few techniques that can test this model, yielding to the measurement of the size and of temperature profile of the accretion disc. We present spectroscopic observations of the gravitationally lensed broad absorption line quasar \obj, which reveal partial microlensing of the continuum emission that appears to originate from two separated regions: a microlensed region, corresponding the compact accretion disc; and a non-microlensed region, more extended and contributing to at least 30\% of the total UV-continuum flux. Because this extended continuum is occulted by the broad absorption line clouds, it is not associated with the host galaxy, but rather with light scattered in the neighbourhood of the central engine. We measure the amplitude of microlensing of the compact continuum over the rest-frame wavelength range 1000-7000\,\AA. Following a Bayesian scheme, we confront our measurements to microlensing simulations of an accretion disc with a temperature varying as $T \propto R^{-1/\nu}$. We find a most likely source half-light radius of $R_{1/2} = 0.61 \times 10^{16}\,$cm (i.e., 0.002\,pc) at 0.18\,$\mu$m, and a most-likely index of $\nu=0.4$. The standard disc ($\nu=4/3$) model is not ruled out by our data, and is found within the 95\% confidence interval associated with our measurements. We demonstrate that, for \obj, the existence of an extended continuum in addition to the disc emission only has a small impact on the inferred disc parameters, and is unlikely to solve the tension between the microlensing source size and standard disc sizes, as previously reported in the literature.}

\titlerunning{Extended continuum emission in \obj.}
\authorrunning{D. Sluse et al.}

   \keywords{Gravitational lensing: micro, strong, quasars: general}

   \maketitle

\section{Introduction}

The central black holes in quasars accrete the matter in their surroundings and emit light over the whole electromagnetic range. The standard model of continuum emission from an accretion disc devised by \cite{Shakura1973} predicts that the temperature of the accretion disc, far from its inner edge, varies with the distance to the black hole as $T \propto R^{-3/4}$. Under the assumption of local blackbody emission, this implies that the emitted flux increases with the frequency as in $f_{\nu} \propto \nu^{1/3}$. The emission from the continuum is sometimes altered in outflowing gas, as revealed by the presence of deep absorption troughs observed in 20-40\% of the quasars. Broad absorption line (BAL) quasars constitute a class of objects, which show absorption in high ionization lines (such as \CIV) with velocities reaching several tens of thousands kilometers per second. The structure and kinematics of this wind is complex, possibly constituted by multiple components \citep[][]{Lamy2004, Borguet2010, Odowd2015}. Variability studies indicate that the wind may be composed of high velocity clouds located at a few tenths of parsecs from the disc \citep{Capellupo2013} in some objects, or at distances of several (hundreds of) parsecs in others \citep{Arav2013}. 

\obj~is a famous BAL quasar at redshift $z_s=2.556$. The light rays originating from this object are bended by a foreground lensing galaxy at a putative redshift $z_{lens} \sim 1.0$ \citep{Magain1988, Kneib1998}, resulting in the observation of four images of this distant quasar. Each of these images is observed through a different portion of the lensing galaxy, such that the stars in the lens act as secondary gravitational microlenses that selectively magnify the source on scales of about 0.01 pc. This scale matches roughly the expected rest-frame ultraviolet size of the accretion disc predicted by the standard model. 

Since 1989, (rest-frame) UV spectra of the individual lensed images have been obtained almost every five years. Those data have revealed that the continuum in image D is about 0.7\,mag brighter than what is measured in the broad and narrow emission lines. This brightening of image D is due to microlensing magnification of the continuum \citep[][ hereafter Paper I]{Angonin1990, Hutsemekers1993, Ostensen1997, Chae2001, Hutsemekers2010}. In 2011, we obtained new spectra of \obj~covering the UV-optical rest-frame domain (1000-7000\,\AA). In the present paper, we use these spectra to study the chromatic change of microlensing. The data also reveal a different amplitude of microlensing between the unabsorbed and the absorbed continuum fluxes, suggesting two sources of continuum: a compact region that is microlensed, and a more extended region that is not (or less) microlensed. In addition, we estimate how this extended continuum modifies the size and temperature profile of the disc, as derived from chromatic microlensing. 

The structure of the paper is as follows. In Sect.~\ref{sec:data}, we present the data. Section~\ref{sec:analysis} explains how we use the microlensing signal from our spectra to unveil the existence of a non-microlensed continuum, and to measure the chromatic dependence of the amplitude of microlensing affecting the compact continuum. In Sect.~\ref{sec:characteristics}, after demonstrating the robustness of our detection of an extended continuum, we characterize its properties and discuss its physical origin. The constraints we infer on the size and temperature profile of the accretion disc are presented in Sect.~\ref{sec:discsize}. 
 
\section{Observations and data reduction}
\label{sec:data}

Our principal data set are spectra obtained in 2011 with the FORS2 and SINFONI instruments mounted on the Very Large Telescope (PI: Hutsem\'ekers, program ID:386.B-0337). These data are described hereafter. We complemented those data with archival optical spectra obtained on 1989-03-07 at the Canada-France-Hawai Telescope, and on 1993-06-23 and 2000-04-21-26 with the Hubble Space Telescope (see Paper\,I for a complete description). 

\subsection{The SINFONI data}
\label{sec:sinfoni}
\obj~has been observed on 2011-03-27 with the SINFONI H+K grating ($R\,\sim\,1500$), dithering the object between four positions on the detector and two different position angles on the sky (0\degr and 90\degr). This resulted in eight exposures of the system, 600\,s each. The sky spectrum was simultaneously obtained on each frame. The seeing was around 0.6\arcsec. Reductions were performed using the Esorex-SINFONI pipeline (v2.2.9). The eight exposures were processed 2 by 2 to obtain four data cubes (wavelength-calibrated and sky-subtracted) with positive and negative images or spectra in each of them. The spectra of the individual components of the quadruple-lensed quasar were extracted by fitting four Gaussian profiles to the positive and negative images of the quasar for each wavelength plane independently. Fitting Moffat profiles gave identical results. Telluric absorption lines were corrected using the spectra of standard stars normalized by a blackbody. After a careful examination of the eight spectra of each quasar image, we selected the four best exposures that provided spectra in perfect agreement. 

\subsection{The FORS2 data}
\label{sec:fors2}

The FORS2 spectra were obtained by positioning a 0\farcs4-wide multi object spectroscopy (MOS) slit through the $A-D$ images, and through the $B-C$ images. The spectra of $B$ and $C$ are ignored in this analysis due to the small amount of microlensing in those images (i.e., about 0.25\,mag). The grism 300V was used to cover the 3600-9000\,\AA\, spectral range. Each observation of $A-D$ was repeated at two epochs approximately separated by the time delay between the two images. The first data set was obtained on 2011-03-05 (hereafter epoch 1), and the second  on 2011-03-31 (epoch 2). The spectra were obtained in the same range of airmasses and parallactic angle at epochs 1 and 2 (i.e., airmass 1.404-1.281 at epoch 1 and 1.398-1.278 at epoch 2) such as they can be compared one-by-one directly without uncertain differential extinction corrections. The observations were carried out in the FORS2 spectropolarimetric mode. Exposures of 660\,s each were secured with the half-wave plate (HWP) rotated at the position angles 0\degr, 22.5\degr, 45\degr,  and 67.5\degr. For each image and grism, 16 spectra (two orthogonal polarizations, four half-wave plate angles HWP, two epochs) were obtained and extracted by fitting two Moffat profiles along the slit. The analysis of the polarization is presented in a separate paper (Hutsem\'ekers et al., in prep).  We corrected the spectra of epoch 2 for slit losses caused by seeing differences to compare the spectra
at epochs 1 and 2 . For that purpose, we used the spectra of three field stars observed simultaneously to \obj. The comparison of the stellar spectra allowed us to verify the absence of chromatic changes at a level larger than 5\% (from 3600 to 9000\AA), and to derive for each position of the HWP a multiplicative correction factor. An average factor epoch2/epoch1 $\simeq$ 0.87 was found with a scatter of $\sim 5\%$. A more significant corrective factor (epoch2/epoch1 $\simeq$ 0.65) was found for the HWP orientation of 67.5\degr due to the poorer seeing of the data obtained at epoch 2. Because of the larger uncertainties affecting that data set, we excluded it from the present analysis. 

\section{Evidence for two continuum sources from the microlensing analysis}
\label{sec:analysis}

\begin{figure*}[!ht]
        \includegraphics[width=1.05\hsize]{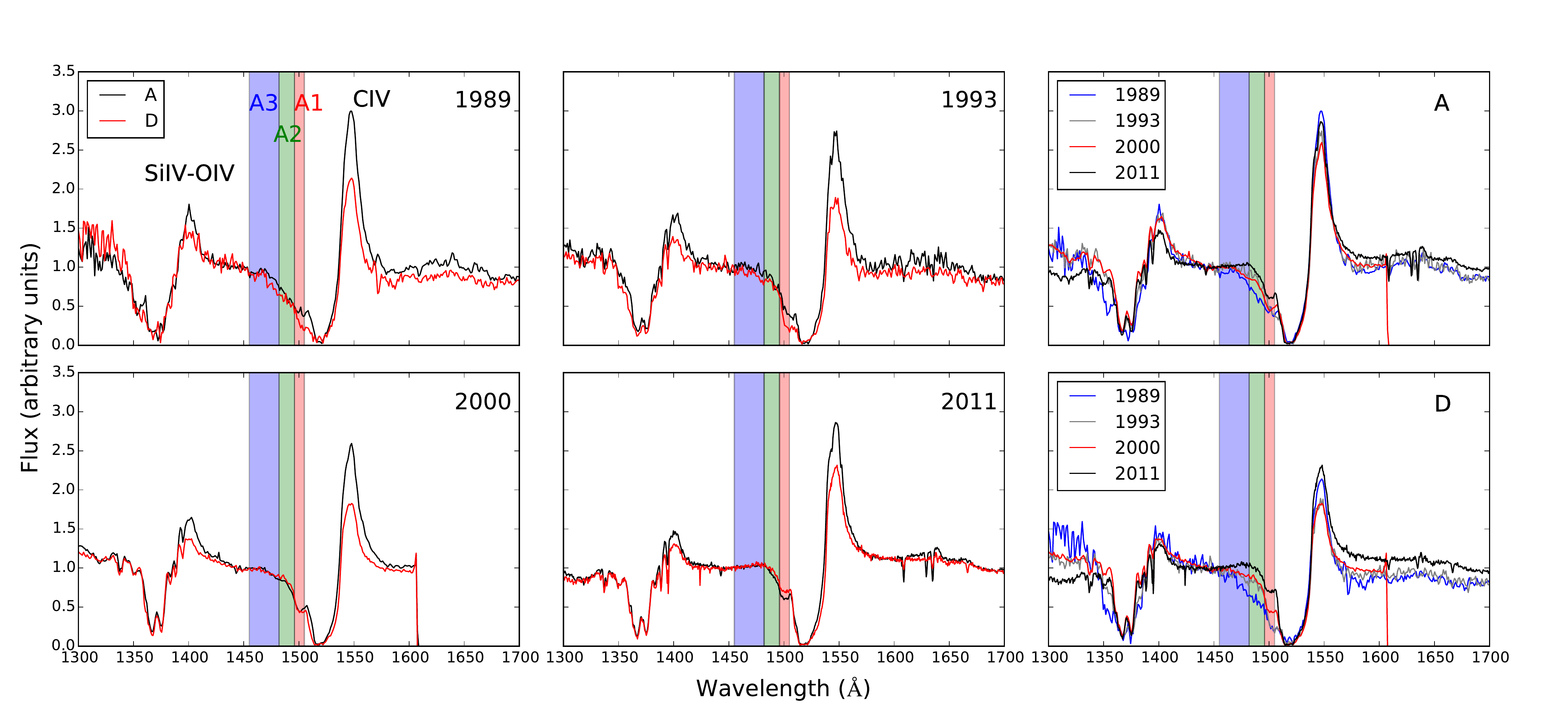} 
        \caption{Spectra of images $A$ (black) and $D$ (red) in the spectral range 1300-1700\,\AA, covering the \SiIV-\CIV\, lines. The left and middle panels show the spectra as observed in 1989, 1993, 2000, and 2011, after normalization to unity in the continuum. The three shaded areas correspond to regions where the absorption changed significantly over the past 20 years, as discussed in Sect.~\ref{sec:microCIV}. The top and bottom right panels show the change of intrinsic absorption over time for images $A$ and $D$.}
        \label{fig:civ}
\end{figure*}

Our analysis focuses on the microlensing in image $D$, which is compared to image $A$. The latter image did not show evidence for significant microlensing in the rest-frame UV wavelength range, over the last ten years (Paper I), and the current data confirm this finding. Although the intrinsic variability of \obj~is relatively small, ambiguity between differences caused by intrinsic variability and microlensing can be present. Fortunately, our new spectra  have been obtained at two epochs separated by 25 days, matching the time delay $\Delta t = 23 \pm 4 $\,days between images $A-D$ \citep{Goicoechea2010, Kumar2014}. Unless otherwise stated, when we present spectra of $A$ and $D$ in the (rest-frame) UV range, we use the spectra separated by the time delay, namely the spectrum of $A$ obtained on 2011-03-05 and that of $D$ obtained on 2011-03-31.

\subsection{Methodology}

We follow two complementary methods to characterize the microlensing signal. On the one hand, the amplitude of microlensing as a function of wavelength is measured using the ``MmD'' decomposition technique \citep[][]{Sluse2007, Hutsemekers2010, Sluse2012a, Braibant2014}, briefly summarized hereafter (Sect.~\ref{MmD}). On the other hand, measurement of microlensing in the \CIV\, absorption trough, is realized by calculating the equivalent width (EW) of the absorption line over small velocity slices (Sect.~\ref{EW}).

\subsubsection{The MmD}
\label{MmD}
Assuming that the observed spectra $F_i$ of the lensed image $i$ is the sum of a spectrum $F_M$, which is only macrolensed, and of a spectrum $F_{M\mu}$, which is both macro- and microlensed, it is possible to extract  components $F_M$ and $F_{M\mu}$  using pairs of observed spectra. Indeed, considering a line profile, we can write

\begin{align} 
F_1  =\, & M F_M + M \mu F_{M\mu} \\ 
F_2  =\, & F_M +F_{M\mu} \; ,
\end{align}

\noindent where $M=M_1/M_2$ is the macroamplification ratio between images~1 and~2 and $\mu$ the microamplification factor of image~1. We assume that image 2 is not affected by microlensing. We can conveniently rewrite these equation to isolate the components of interest: 

\begin{eqnarray} 
F_M \ & = & \frac{-A \;}{A - M} \; \; \left( \frac{F_1}{A} - F_2 \right), 
\\ 
F_{M\mu}
& = & \frac{M}{A - M } \; \; \left( \frac{F_1}{M} - F_2 \right),  
\end{eqnarray} 

\noindent where $A = M\mu$. Up to a scaling factor, $F_M$ only depends on $A$, while $F_{M\mu}$ only depends on $M$. $A$ is the scaling factor between the $F_1$ and $F_2$ continua. It can be accurately determined as the value for which $F_M(A) = 0$ in the continuum. The microamplification factor of the continuum is then estimated using $\mu = A/M$. 

For the MmD of images $A$ and $D$ presented hereafter, we fix $M=0.40\pm0.02$ prior to the decomposition (instead of the empirical determination performed in earlier papers). This value is consistent with values retrieved around $\Hbeta + \OIII$ with SINFONI data (Paper~I, $M=0.38\pm0.01$), and with the mid-IR flux ratios obtained by Mac Leod et al. (\citeyear{McLeod2009}; $M=0.40\pm0.06$).

\subsubsection{The equivalent width}
\label{EW}

The EW, because it involves a normalization by the continuum, is a simple method to unveil differential microlensing between the continuum and another source of emission. By definition, the equivalent width of a line (in emission or absorption) is 
\begin{equation} 
EW = \int_{\lambda_1}^{\lambda_2} \left(1-\frac{F_\lambda}{\mathcal{F}}\right) \,d\lambda,
\label{equ:EW}
\end{equation}
\noindent where $\mathcal{F}$ is the continuum intensity underlying the absorption or emission line and usually approximated by the continuum emission on either side of the studied feature. The flux over the wavelength range of interest $[\lambda_1, \lambda_2]$ is denoted $F_\lambda$. It is easy to see that the EW is only  modified due to microlensing  if $F_\lambda$ and $\mathcal{F}$ are microlensed at different levels, i.e., if the microlensed fluxes are $\mathcal{F}^{\prime} = \mu_c \mathcal{F}$, and $F^{\prime}_\lambda = \mu_\lambda F_\lambda$, the EW only changes  if  $\mu_\lambda \neq \mu_c$.

\subsection{Microlensing around the \CIV\, line}
\label{sec:microCIV}

First, we focus on the region of the \CIVd\,line. Figure~\ref{fig:civ} shows the variations observed in images $A$ and $D$ between 1989 and 2011. The visual inspection of the spectra, normalized in the spectral range 1445-1455\,\AA, free of line emission, shows that the two spectra do not match, and,  in particular, the flux in the emission lines (e.g., the peaks of the \CIV\, and \SiIV-\OIV) is smaller in image $D$ than in $A$. This has been observed previously and is caused by microlensing amplification of the continuum of image $D$. Indeed, because we have normalized the spectra in the continuum region, the magnification of the continuum of $D$ leads to an apparently less intense emission line in that image, because the line is actually unaffected by microlensing.

\begin{figure}[tb]
        \centering
        \includegraphics[width=1.05\hsize]{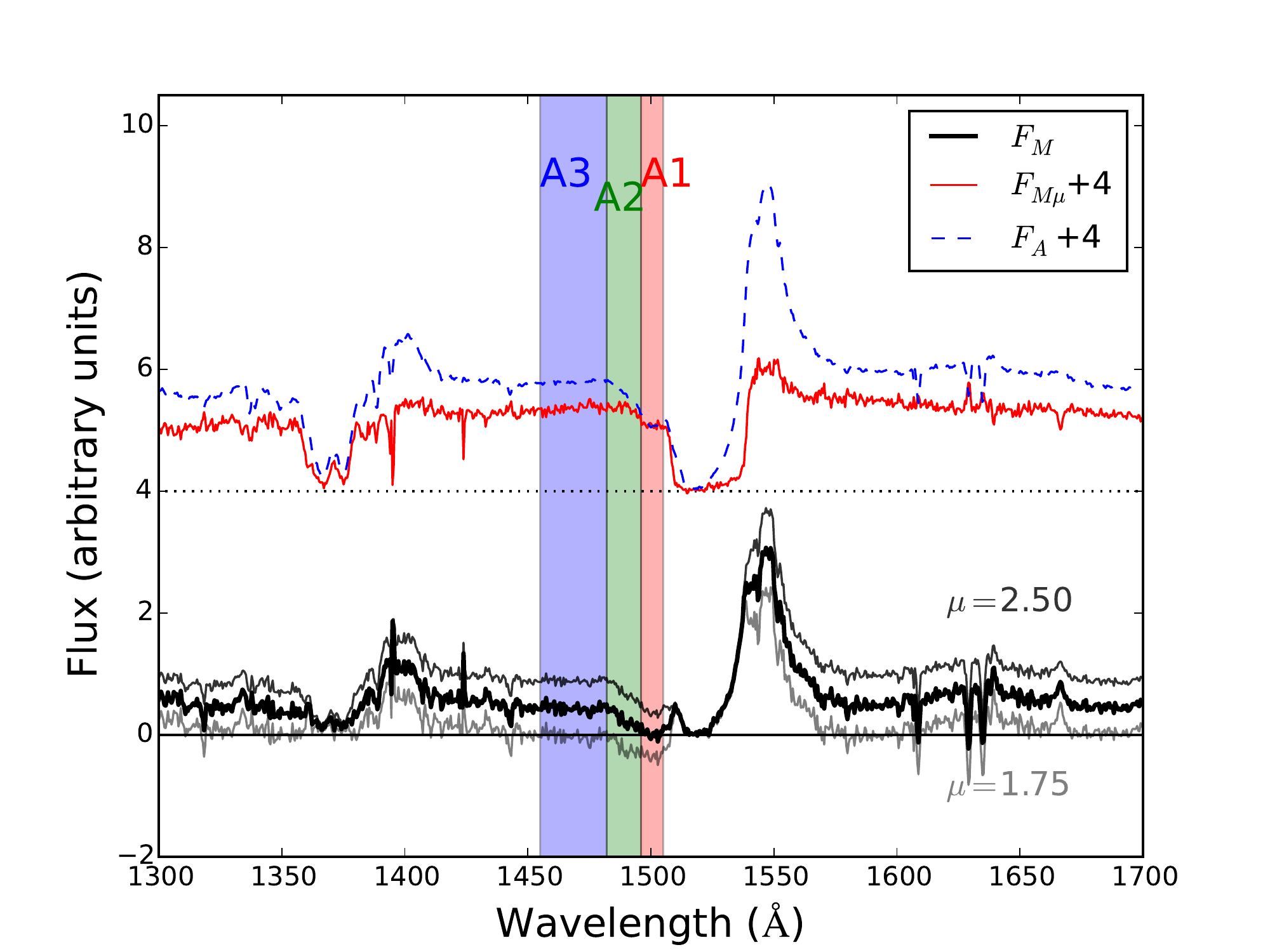}        
        \caption{MmD performed on images $A$ and $D$ of \obj\,in 2011, using $M=0.4$ (Sect.~\ref{sec:chroma}). The dashed blue line is the flux of image $A$. The thick black line is the fraction of the flux $F_M$ unaffected by microlensing for $\mu=2.0$. The red line is the corresponding microlensed fraction of the flux $F_{M\mu}$ (shifted up to ease legibility), the dotted horizontal line is the line corresponding to $F_{M\mu} =0$. The gray lines show the decomposition for two other values of $\mu$ (see Sect.~\ref{sec:microCIV}). The three shaded areas correspond to the three regions $A1-A2-A3$ as discussed in the text.}
        \label{fig:MmDciv}
\end{figure}

The other important feature to notice in Fig.~\ref{fig:civ}, is the difference of absorption depth in $A$ and $D$ on the blue side of  \CIV. This is particularly noticeable in the range 1455-1505\,\AA, delimited with three shaded areas, and where the absorption is less deep. On the one hand, we see a decrease of the depth of the absorption with time. This corresponds to a long-term intrinsic decrease of the absorption. On the other hand, we observe a relative change of the absorption in $A$ and $D$ over time: prior to 2000, the absorption is deeper in image $D$ than in $A$; in 2000, it is similar in $A$ and $D$; and finally in 2011, it is deeper in image $A$. Because of this inversion of the relative absorption between $A$ and $D$ in 2011, {\it \textit{\textup{at that epoch it is impossible to perform  a MmD with a zero continuum everywhere} \textup{in}} $F_M$}, as shown in Fig.~\ref{fig:MmDciv}. Indeed, when we enforce the continuum to be zero in $F_M$, setting $\mu=1.75$, we violate the positivity constraint on $F_M$ at the level of the absorber in the blue side of the \CIV\,line (see  light gray line in Fig.~\ref{fig:MmDciv}). In order to get $F_M = 0$ in that range, it is necessary to allow for a larger amount of microlensing of the continuum: $\mu=2.0$ ($F_M$ shown with a thick black line in Fig.~\ref{fig:MmDciv}). This new value of $\mu$ yields $F_M > 0$ in the unabsorbed continuum flux. At the same time, a large fraction of the continuum flux is still visible in $F_{M\mu}$ (i.e., the continuum emission above the dotted line in Fig.~\ref{fig:MmDciv}). Therefore, the MmD suggests that there are two components producing the continuum, one originating from a region, which is sufficiently compact to be microlensed, visible in $F_{M\mu}$, and another one that we detect in $F_M$, and which comes from a nonmicrolensed, and, hence, more extended region. We emphasize that this detection of an extended continuum emission with the MmD also implies that the compact and the extended continua are differently absorbed. More specifically, the extended continuum has to be {\it{\textup{more absorbed}}} than the compact continuum (in the range 1455-1505\,\AA). As explained in Appendix~\ref{AppendixMmD}, if both continua were similarly absorbed by the outflowing gas, or if the most compact region was more absorbed than the extended region, the flux from the latter would remain undetected with the MmD when imposing $F_M = 0$ at the wavelength of the \CIV\, absorber (i.e., $\lambda \sim$ 1500\,\AA). On the other hand, we may not exclude larger magnification of the compact continuum. Fig.~\ref{fig:MmDciv} shows the decomposition for $\mu=2.5$ (top gray line). In that case, the intensity of the extended continuum emission equals the intensity of the compact (microlensed) continuum.

\begin{figure}  
        \includegraphics[width=\hsize]{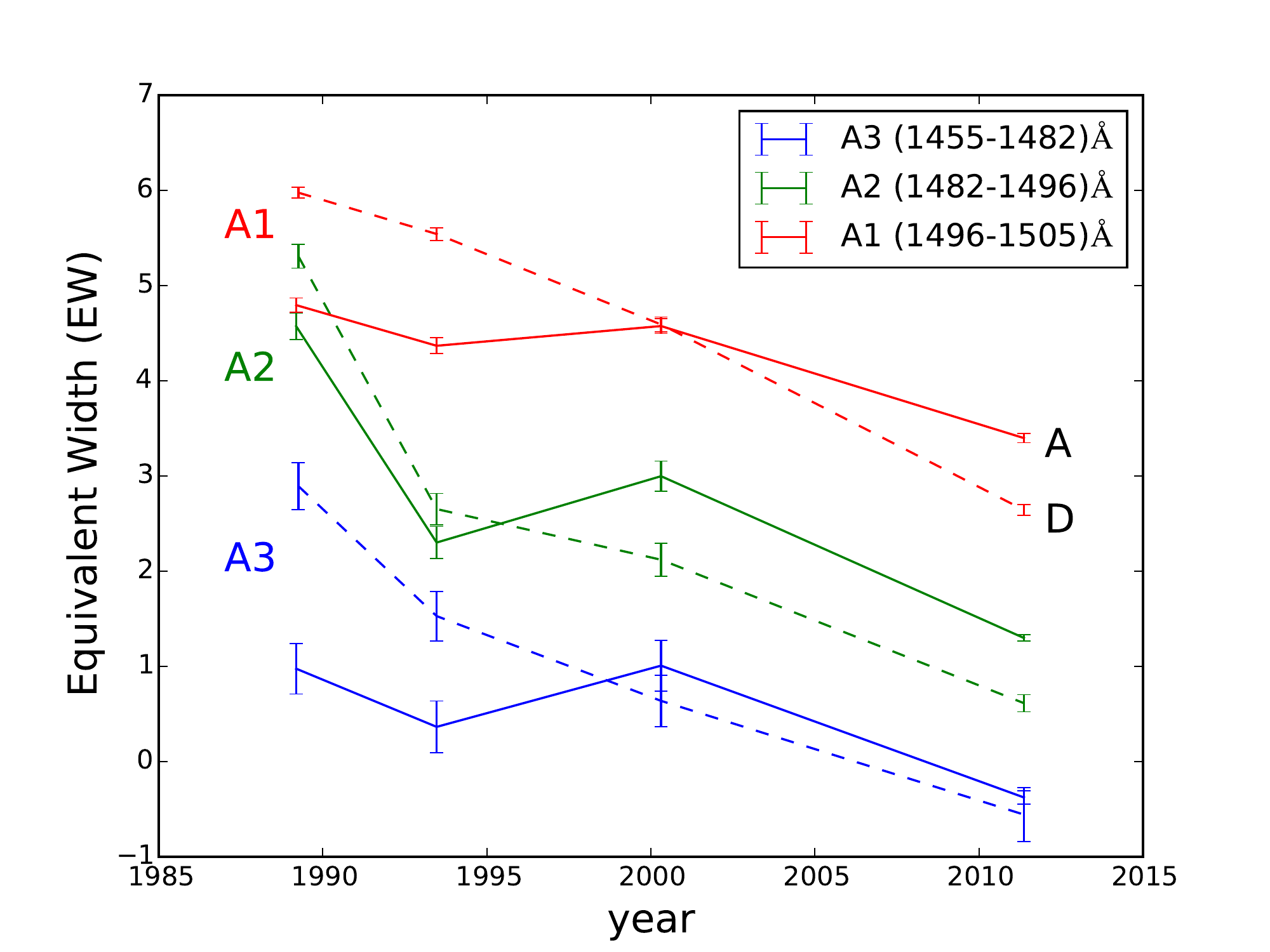}
        \caption{Time variation of the equivalent width of the absorption, in the wavelength ranges $A1$-$A2$-$A3$ (see legend), for image $A$ (solid) and $D$ (dashed).} 
        
        \label{fig:ew}
\end{figure}

To better characterize this differential signal in $A$ and $D$, we show in Fig.~\ref{fig:ew} the time variation of the EW of the absorption in three different wavelength ranges (Fig.~\ref{fig:civ} and \ref{fig:MmDciv}): the range $A1=[$1496, 1505$]$\,\AA\,([-10274, -8532]\,km/s), is a flat absorption plateau, which barely changes profile over time, but whose depth is significantly decreasing; the range $A2=[$1482, 1496$]$\,\AA\,([-12984, -10274]\,km/s) covers the highest velocity part of the absorption as observed in 2011; and finally  $A3=[$1455, 1482$]$\,\AA\,([-18209, -12984]\,km/s) is the range where absorption was detected prior to 2011. The variation of EW, shown in Fig.~\ref{fig:ew}, allows us to assess the significance of the relative absorption differences detected in Fig.~\ref{fig:civ}. The EW declines over the 22 years time span of the observations, in agreement with the intrinsic decrease of the absorption depth seen in Fig.~\ref{fig:civ}. In addition, while $EW_A < EW_D$ before 2000, the reverse behavior is detected in 2011. The observed differences of EW in $A1$ and $A2$ are highly significant. We demonstrate in Appendix~\ref{AppendixEW} that it is only possible to observe $EW_A > EW_D$ in presence of two sources of continuum light, which are differently absorbed {\it{\textup{and}}} microlensed. The difference of EW observed between $A$ and $D$ at other epochs can also be understood under this hypothesis, but in that case the interpretation is not unique, as $EW_A < EW_D$ can also be explained by superimposed broad line emission at the wavelength of the absorber. 

To summarize, the MmD decomposition and the EW variation reveal the existence of two sources of continuum, a compact continuum that is microlensed, and a continuum that is more extended, and hence not microlensed. The MmD reveals the extended continuum in a visual and intuitive way, while the EW allows us to assess the significance of the observed signal. 

\subsection{Microlensing around other lines: chromatic microlensing}
\label{sec:chroma}

\begin{figure*}
\begin{tabular}{cc}
          \includegraphics[width=0.5\hsize]{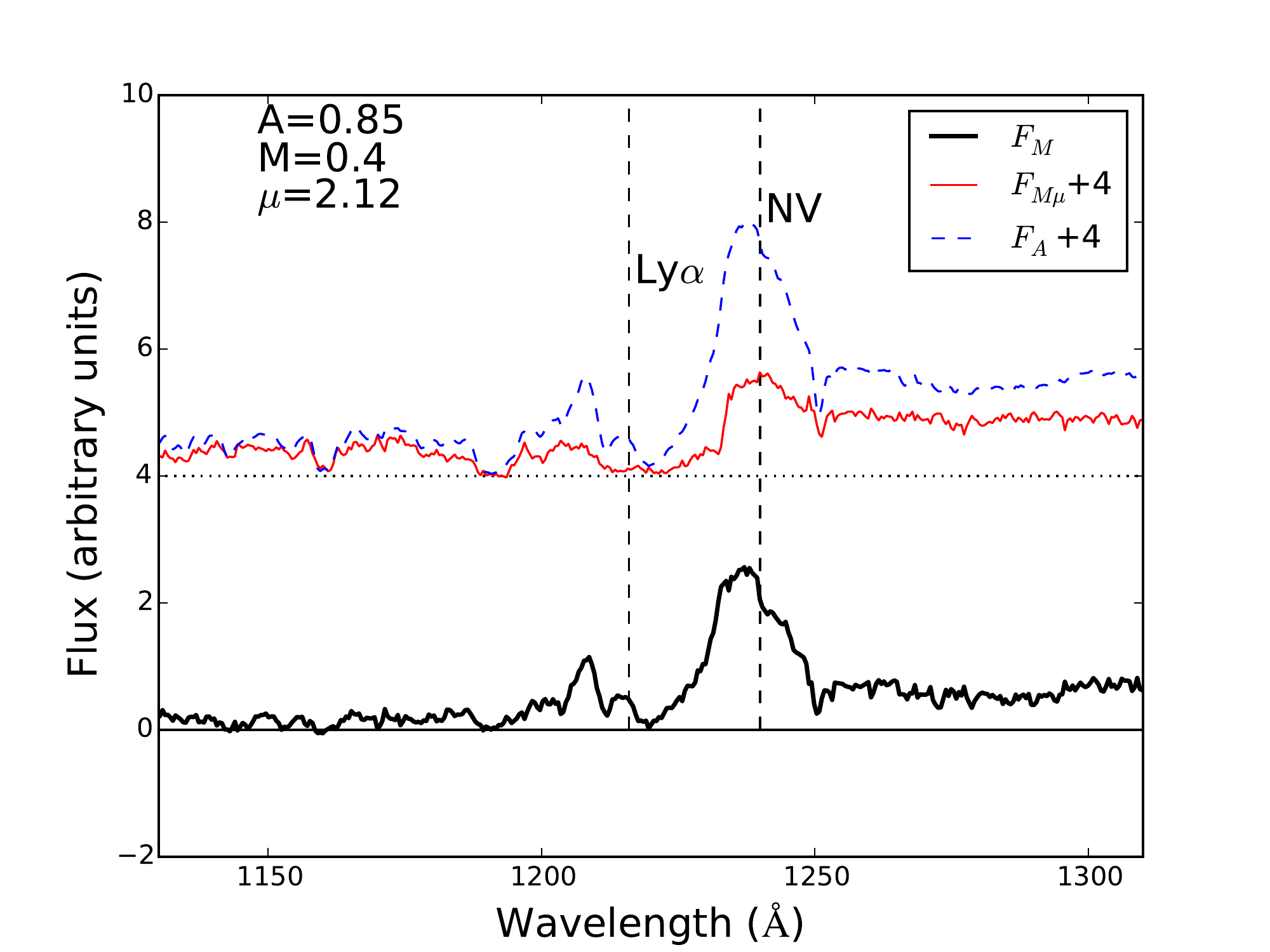} & \includegraphics[width=0.5\hsize]{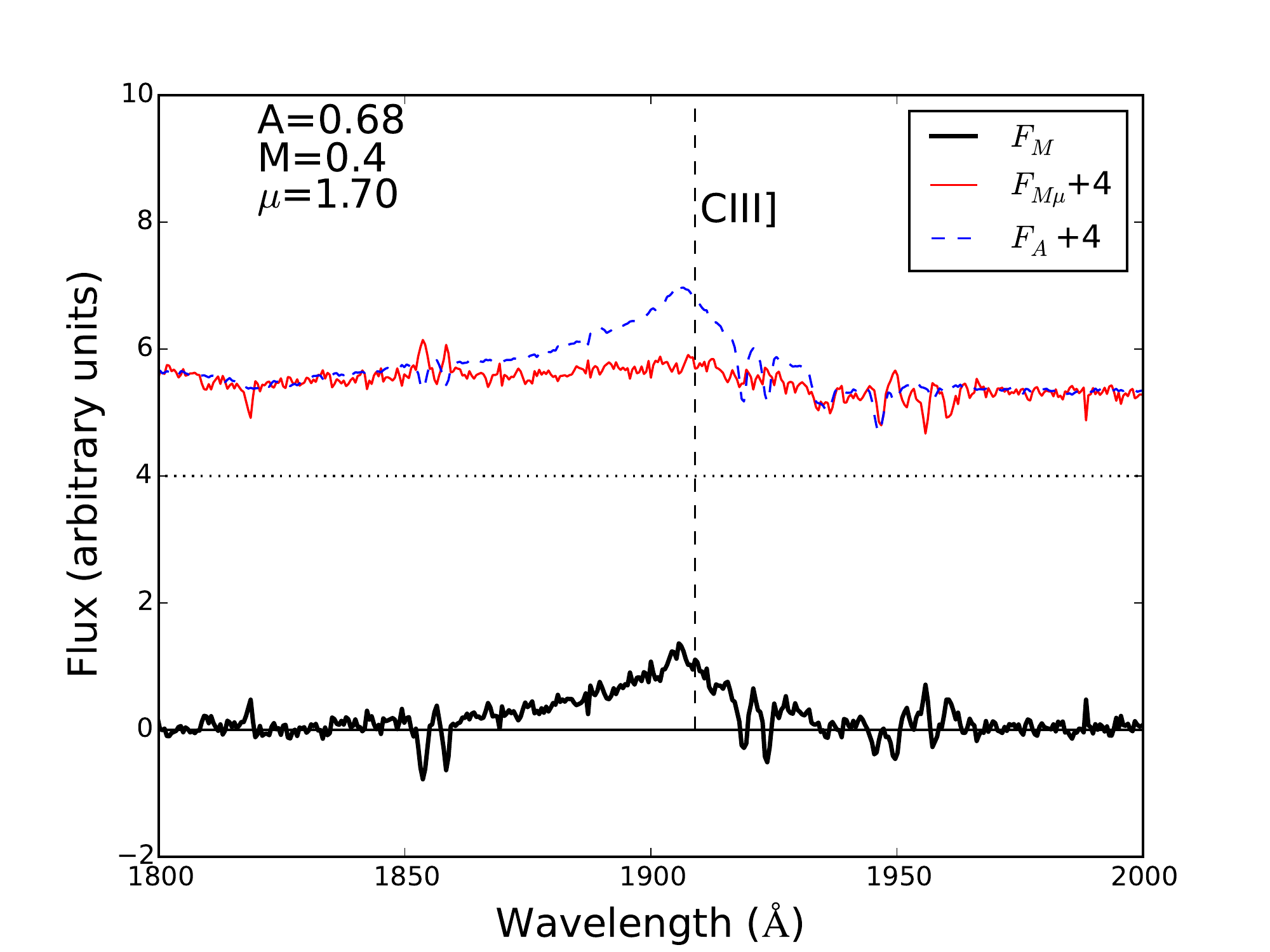} \\
          \includegraphics[width=0.5\hsize]{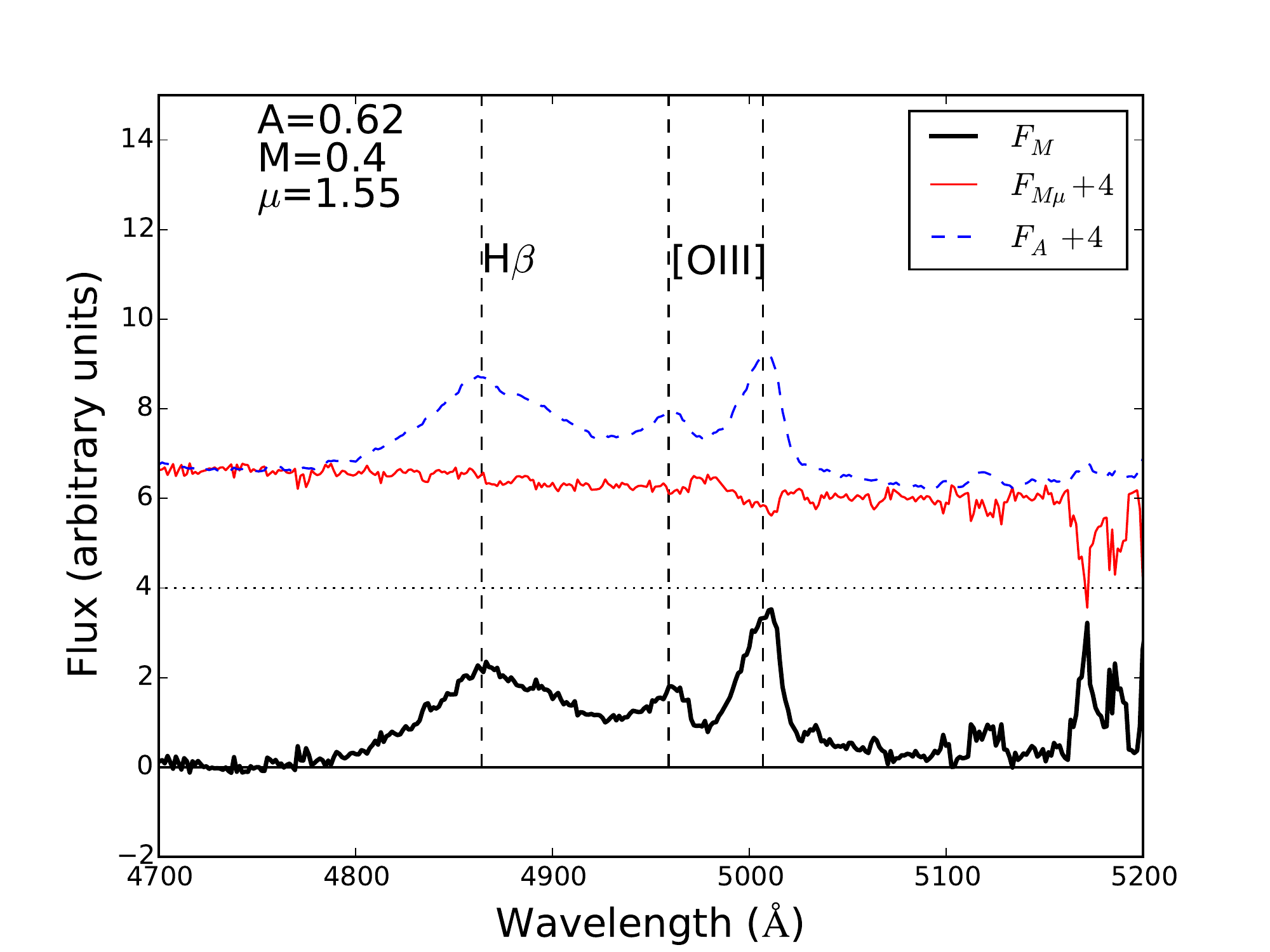} & \includegraphics[width=0.5\hsize]{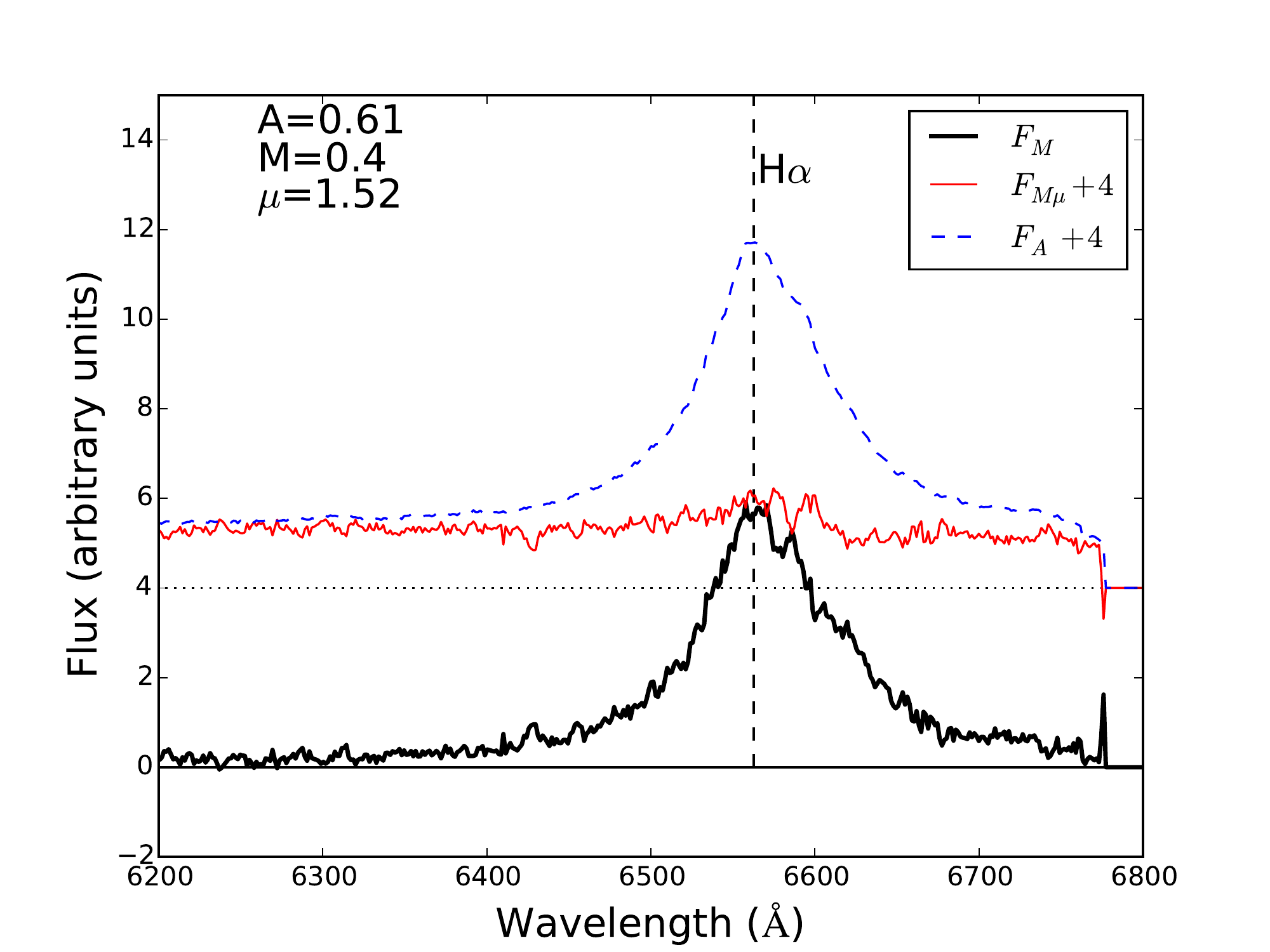} \\
\end{tabular}
        \caption{MmD for \Lyalpha\,+\NV\,(top left), \CIII\,(top right), \Hbeta\,(bottom left), \Halpha\,(bottom right) emissions.  The values of $A = M \times \mu$ derived for each line are based on the individual frames, while the MmD is shown for the sum of the individual spectra.   } 
        \label{fig:MmD}
\end{figure*}

The spectra obtained in 2011 cover a large fraction of the wavelength range [1100, 7000]\,\AA\, (rest frame). Using the  MmD technique (Sect.~\ref{MmD}), we can derive the amplitude of microlensing as a function of wavelength. Because of the lack of apparent chromaticity of $M$ empirically measured in past optical and NIR spectra (Paper I), we may assume that there is no significant differential extinction between $A$ and $D${\footnote{\cite{Munoz2011} suggests possible extinction in $A$, but the time variability of the various flux ratios does not allow us to assess the uniqueness of this interpretation. As pointed out by \cite{Munoz2011}, extinction estimates are very uncertain in \obj. However, extinction in $A$ is smaller than in $B$, hence reducing the observed flux of $A$ by less than 10\% (cf. Paper I for extinction in $B$). This has a negligible impact on our source size estimates (Sect.~\ref{sec:discsize}).}}. The measurement of the amplitude of microlensing, in the vicinity of the main emission lines, is repeated for each individual spectra. The standard deviation between measurements (from three spectra for FORS2 data and four spectra for SINFONI) is quadratically added to the uncertainty in the relative flux calibration between epochs (UV data), and used to calculate the standard error on the mean. The results of this procedure are summarized in Table~\ref{tab:MmD-DA}. The MmD of the main emission lines is presented in Fig.~\ref{fig:MmD}. As explained above, the measurements in the rest-frame UV account for the time delay. 
Although intrinsic variability of the absorption lines is noticeable between the two epochs, it mostly affects the value of $\mu$ derived from \CIV. In that case, measurements of $\mu$ on spectra obtained at a single epoch nevertheless agree at the 3$\sigma$ level with those presented in Table~\ref{tab:MmD-DA}. We may then be confident that the values of $\mu$ derived at a single epoch from the H$\beta$ and H$\alpha$ lines are accurate, in particular, since they are identical within the errors to those derived in Paper I.  

The extended continuum detected around \CIV\,(Sect.~\ref{sec:microCIV}) is also tentatively detected around \Lyalpha\,-\NV. In that case, the evidence for an extended continuum flux is set by the absorbers in the range [1170-1190]\,\AA\,(rest frame), which have been identified as \ion{H}{i} absorption at $z\sim 2.4-2.45$ \citep[line number 48 to 51 in ][]{Monier1998}. They are therefore likely to be high velocity absorptions associated with \obj. Note that in the MmD of the \Lyalpha\,line shown in Fig.~\ref{fig:MmD}, the positivity constraint in $F_M$ is (marginally) violated around $\lambda \sim 1165\,$\AA. This is acceptable because this absorption feature is due to intervening \CIV\,\citep[line number 45, 46 in ][]{Monier1998}, and therefore may be intrinsically different in $A$ and $D$. On the other hand, the absorption in the blue side of \OIV-\SiIV\, sets almost no constraint on the presence of an extended continuum in that wavelength range. This is because the well-identified high velocity absorption $A1-A2$, visible in \CIV, is shallower and less clearly detected, in the  \OIV-\SiIV\, blend. 

\subsection{Differences of light paths between images}
\label{sec:lightpaths}

Up to now, we have shown that the continuum flux from \obj\, originates from two regions, a compact one, which is small enough to be microlensed, and a more extended one, which is not microlensed. An alternative interpretation could be that significant inhomogeneities of the outflowing absorbing material lead to different absorption for each of the lensed images. This can be tested by estimating the linear comoving separation between the two different lensed images $A$ and $D$ as a function of the distance to the central black hole. The comoving linear beam separation for $z > z_{lens}$~\citep{Smette1992, Chartas2007} is given by
\begin{equation}
S(z_a) = \theta \frac{D_{ol} D_{sa}}{D_{sl}},
\end{equation}

\noindent where $\theta$ is the separation between the lensed images, $D$ is the angular diameter distance, and the subscripts $o$, $l$, $s$, $a$ refer to the observer, lens, source, and absorber, respectively. 

In order to see if the comoving distance between light paths is larger than a cloud size, we need to assume a reasonable cloud distance to the central engine. Important constraints on the distance and geometry of absorbing troughs have been set by variability studies of BALs. Two scenarios are generally considered. One explaining the variability as caused by rapid crossing of an absorbing cloud, and one as a rapid change of ionization in the ionized material \citep[e.g.,][]{Lundgren2007, FilizAk2012, Capellupo2013, Wildy2015, Grier2015}. Recently, \cite{Capellupo2014}  carried out a detailed study of the variability of the A1-A2 absorbers in \obj. Based on the comparison of the \CIV\, and \PV\, absorptions, they showed that the \CIV\, absorption is saturated such that its variability is most likely caused by rapid crossing of an absorbing cloud. Assuming a so-called crossing-disc model, they derive an upper limit of 3.5\,pc on the radial distance of the cloud to the black hole. At this distance, the separation between $A$ and $D$ corresponds to a linear beam separation of $ 3.2 \times 10^{13}$\,cm, ($10^{-5}\,$pc) hence about ten times smaller than the gravitational radius of the source, ensuring that the two lines of sight cross the same absorber. Therefore, the possibility that the light from image $A$ and $D$ intercept different clouds (or regions of a cloud with different opacities) is very unlikely.

\begin{table}[t]
        \caption{Amplification factors for the (D,A) pair, with fixed $M$.}
        \label{tab:MmD-DA}
        \begin{tabular}{lcccc}\hline\hline \\[-0.10in]
                Lines &  $\lambda$ (\AA)&  $A$ & $M$ & $\mu$\\ 
                \hline \\[-0.10in]
                \Lyalpha+\NV & 1216 & 0.85$\pm$0.01  & 0.40$\pm$0.02 & 2.12$\pm$0.11 \\
                \SiIV+\OIV           & 1402 & 0.70$\pm$0.01 & 0.40$\pm$0.02 & 1.74$\pm$0.09 \\
                \CIV           & 1549 & 0.80$\pm$0.01  & 0.40$\pm$0.02 & 2.00$\pm$0.10\\
                \CIII         & 1909 & 0.68$\pm$0.01  & 0.40$\pm$0.02 & 1.74$\pm$0.09 \\
                \Hbeta      & 4861 & 0.62$\pm$0.01  & 0.40$\pm$0.02 & 1.55$\pm$0.08$\;$ \\
                \Halpha     & 6563 & 0.62$\pm$0.01  & 0.40$\pm$0.02 & 1.54$\pm$0.08$\;$ \\
                
                \hline\\[-0.2cm]
        \end{tabular}\\
        {\small {The measurements for UV lines are performed on pairs of spectra separated by the time delay (i.e., 2011-03-05 for $A$ and 2011-03-31 for $D$). The NIR measurements are performed on single epoch spectra obtained on 2011-03-27. For \Lyalpha+\NV\,(resp. \CIV), the values are obtained for $F_M=0$ in the range [1170-1190]\AA\ (resp. [1496-1505]\AA). For \SiIV+\OIV\, and \CIII\,, the value of $\mu$ is likely underestimated because of the unknown contribution of the extended continuum at those wavelengths.}}
\end{table}

\section{Characteristics of the extended continuum}
\label{sec:characteristics}

In this section, we first try to characterize some of the properties of the extended continuum, in particular its wavelength dependence, and its possible time variability. We then discuss its origin.

\subsection {Wavelength dependence}
\label{subsec:chroma-extended}

The MmD yields the measurement of the relative fractions of compact ($cc$) and extended ($ce$) continua from the total flux. If we define $f$ as the fraction of the total continuum flux $\mathcal{F} = F_{cc}+F_{ce}$, which is extended (i.e., $f=F_{ce}/\mathcal{F}$), we find based on the MmD around \CIV\, and \Lyalpha\, that $f=0.3$ (i.e., 30\% of the total flux is nonmicrolensed). This value of $f$ implies that  $F_{ce}/F_{cc}= f/(1-f)= 0.42$. Larger values of $f$ may be acceptable (Sect.~\ref{sec:microCIV}, Fig.~\ref{fig:MmDciv}). 

Although we do not observe a significant difference of the amount of extended continuum between the \Lyalpha\, and \CIV\, wavelengths, we cannot exclude a small chromatic change. Since the \Lyalpha\, and \CIV\, absorbers used to derive $\mu$ correspond to clouds at different velocities, the opacity of the absorber (i.e., $\tau_{cc}$ and $\tau_{ce}$, Appendix~\ref{AppendixMmD}) may be different in \Lyalpha\, and \CIV, leading to a coincidentally similar measure of $\mu$ around those two lines, but the effect is unlikely to be large. At wavelengths redder than \CIV\,, the situation is more complex as the lack of intrinsic absorption in that range precludes the detection of the extended continuum. At the wavelength of the Balmer lines in the NIR, we measure $\mu \sim 1.5$, both in 2005 and 2011.  This constancy of the amplitude of microlensing suggests that the contribution from a nonmicrolensed continuum is negligible in the redder part of the spectrum (i.e. ,$\lambda > 4800\,$\AA) because the extended continuum
is stronger in 2011 than in 2005 (Sect.~\ref{sec:time}). Consequently, we consider in the following analysis that there is no contribution of the extended continuum at those wavelengths. 

\subsection{Temporal variation}
\label{sec:time}

Our detection of an extended continuum in 2011 has been possible owing to the microlensing of image $D$ and to the differential absorption of the compact and extended continua in 2011, the extended continuum being more absorbed than the compact continuum. We may wonder whether the extended continuum was present at earlier epochs, or simply could not be detected because of the time variation of the relative absorption,  an insufficient signal-to-noise ratio in past data, and/or changes of the microlensing configuration. Actually, the MmD presented in Fig.~6 of Paper I shows that $F_M < 0$ in part of the $A2-A3$ absorption for the data obtained in 2000. Our Fig.~\ref{fig:ew} also reveals that $EW_A > EW_D$ for $A2$ and $A3$ in 2000, confirming (at least from $A2$) a significant detection of the extended continuum at that epoch. Accounting for this absorption in the MmD (i.e., enforcing $F_M = 0$ at the location of the absorber) leads to $F_{ce}/F_{cc} \sim 20\%$ and to $\mu \sim 2.0\pm0.12$. Although the detection is more noisy, this suggests that the extended continuum was indeed present in those data but with a smaller contribution. At earlier epochs, we also detect differences of EW between $A$ and $D$ but with $EW_D > EW_A$. Although this is not an unambiguous signature of an extended continuum since emission from the BLR can also explain this signal (see Appendix~\ref{AppendixEW}), it is plausible to attribute the origin of this difference of EW to the extended continuum. Contrary to the situation observed in 2000 and 2011, the extended continuum has to be less absorbed than the compact one (Appendix~\ref{AppendixEW}). Because we do not know the optical depth of the gas covering the compact and extended continua, we cannot derive $F_{ce}/F_{cc}$ with the MmD or based on the EW prior to 2000. 

Another way to approach the issue of time variability of $F_{ce}/F_{cc}$ is to look at the variation of the effective amplitude of microlensing over time. When calculating $\mu^{\prime}$ determined by making $F_M = 0$ in the unabsorbed continuum adjacent to \CIV\,(see Appendix~\ref{AppendixMmD}), we find it to be roughly constant in 1993/2000/2005 ($\mu^{\prime} \simeq 2.0 \pm 0.2$, Paper~I). In 2011, $\mu^\prime$ significantly drops to 1.75 $\pm$ 0.07. {\it {\textup{Under the hypothesis that the amplitude of microlensing does not significantly change with time}}}, this indicates that a decrease of $\mu^\prime$ results from an increase of $F_{ce}/F_{cc}$ (i.e., Eq~\ref{eq:fracCC}). When we only measure the amplitude of microlensing for the compact continuum  (i.e., allowing $F_M \neq 0$, cf. Appendix~\ref{AppendixMmD}), we find $\mu = 2.0 \pm 0.1$ in 2011 (Table~\ref{tab:MmD-DA}), which is consistent with the amplitude of microlensing in 1993/2000/2005. Moreover, the micromagnification $\mu$ of the \Hbeta\, continuum did not change between 2005 and 2011, also supporting the constancy of $\mu$ in the last decade. This absence of variation of the microlensing signal agrees with the small relative transverse velocity of the lens in \obj~\citep{Mosquera2011}. 

In summary, we find evidence for extended continuum in the past data, but the data also suggest that the fraction of extended over compact continuum varies with time.

\subsection{A physical model}
\label{Twocontinua}

The extended continuum unveiled by our data may simply be genuine flux from the host galaxy, or scattered emission by the host, similar to that unveiled by \cite{Borguet2008} in Type 1, and by \cite{Zakamska2006} in Type 2 AGNs. This interpretation is however very unlikely. Indeed, we have shown that at least 30\% of the total continuum emission is extended. However, \obj\, is a BAL quasar, and most of the light is occulted by the optically thick absorption in the blue wing of \CIV, with no more than a few percent of the flux  detected at those wavelengths. Although this absorber is at least at the distance of the BLR (because it occults the \CIV\, emission), it is unlikely to cover the whole host galaxy. 

Broad absorption line quasars like \obj\, are known to be intrinsically polarized because of scattering of the continuum flux in an extended region possibly located at a radius comparable to the BLR radius or in a polar wind \citep{Lamy2004, Schmidt1999}. It is legitimate to postulate that the same region is at the origin of the polarization {\it {\textup{and}}} the source of extended continuum, especially since the observed fraction of extended continuum is consistent with the polarization degree observed in \obj\,(Hutsem\'ekers et al., in prep.). The scattering could be caused by dust, or by electrons. If dust is responsible for the scattering, we expect a wavelength dependence of the extended continuum as $\lambda^{-4}$. This is consistent with the apparent absence of extended continuum in \Hbeta, but implies an extended continuum emission 2.7 times stronger at the level of \Lyalpha\, than around \CIV. Another possibility is scattering by electrons. This does not yield a chromatic dependence of the extended continuum, in agreement with our observation in the blue. Under this scenario, it remains to be seen how we can explain the lack of extended continuum at the reddest wavelengths, e.g., around \Hbeta\, and \Halpha. A combination of both dust and electron scattering is also possible.

Alternatively, to explain the UV peak in quasar spectral energy distribution (and other inconsistencies between the standard disc model and the SED of quasars), \cite{Lawrence2012} has proposed the existence of a population of dense clouds, which partly reprocesses the intrinsic continuum from the accretion disc. A false continuum is produced in the UV coming from an extended source. Since these clouds mostly reflect the inner continuum through electron scattering, we expect the ratio of the two continua to be roughly constant in the \Lyalpha-\CIV\,spectral range, in agreement with our observations. This model also predicts a decrease of the fraction of extended continuum at reddest wavelength, compatible with the absence of significant extended continuum around $\Hbeta$. In addition, in that model, UV variability is interpreted as changes of the cloud covering factor. It can occur independently of variations of the intrinsic continuum and can explain the change of $F_{ce}/F_{cc}$ between 2000 to 2011 (Sect.~\ref{sec:time}). In this model, the pseudocontinuum flux is suspected to originate from cold clouds located at about $30\,R_{\rm{Sch}}$ (Schwarchild radii) from the continuum, which corresponds to a distance of $0.037$\,pc $= 1.16\times 10^{16}$\,cm $\sim 0.6\,\eta_0$ (for $\log(M_{BH}/M_\odot)=9.12$, and $\avg{M}=0.3\,M_{\odot}$ in Eq.~\ref{equ:RE}). Even if the clouds were located in the range $30-100\,R_{\rm {Sch}}$, the extended pseudocontinuum would be of the order of 1-2\,$\eta_0$, and would therefore be sensibly microlensed. 

In conclusion, we can rule out the host galaxy as  the source of the extended continuum. Instead, we postulate that the latter is scattered light by dust or electrons. Scattering by dust or electrons could take place in the same region as that at the origin of the polarization in BAL quasars. Dust scattering yields a chromatic dependence, while electron scattering does not. Therefore, dust may explain the apparently small amount of extended continuum at the wavelength of \Hbeta\,, but implies that we underestimate the contribution of the extended continuum at the level of \Lyalpha\, by almost a factor 3. Variations of microlensing and of the absorption in front of \CIV\, and \Lyalpha\, may possibly help to test this scenario. Simple electron scattering fails to explain the decrease of extended continuum at redder wavelengths. The model of blurred reflection of the continuum proposed by \cite{Lawrence2012} accounts for most of the observational properties of the extended continuum we detect, except that the cold clouds responsible for the "continuum-reflection" are expected to be located too close to the black hole to remain unaffected by microlensing. 

\section{Constraints on the accretion disc parameters}
\label{sec:discsize}

We now use our measurements of the chromatic microlensing (Sect.~\ref{sec:chroma}) to derive the size of the compact continuum (which originates from the accretion disc) as a function of wavelength. Following various other works \cite[e.g.,][]{Eigenbrod2008b, Bate2008, Anguita2008a}, we test an accretion disc model with a temperature profile $T \propto R^{-1/\nu}$. This kind of model leads to a dependence of the source size $\sigma$ as a function of wavelength $\sigma \propto \lambda^\nu$. This model has two parameters, the source size $\sigma$ (converted into the half-light radius $R_{1/2}$) and the power-law index $\nu$, for which we derive the probability density function (PDF), following the Bayesian scheme described hereafter (Sect.~\ref{Bayes}). 

\subsection{Methodology}
\label{Bayes}

To turn the chromatic variation of microlensing into a quantitative information on the disc, we need to compare this signal to microlensing simulations calculated for different sizes and temperature profiles of the accretion disc, as well as for different fractions of compact objects in the lensing galaxy. The data to which the microlensing simulations are compared are the measurements presented in Table~\ref{tab:MmD-DA}, excluding $\mu$ measured from \CIII\, and \SiIV\, due to the uncertainty in the fraction of extended continuum at those wavelengths. It is important to keep in mind that those values of $\mu$ are derived assuming that the fraction of flux from the extended continuum equals the minimum value needed to reproduce our data (i.e. ,$F_{ce}/F_{cc} = 0.42$, Sect.~\ref{subsec:chroma-extended}).

We follow a Bayesian scheme similar to that developed in \cite{Bate2008}. Throughout the analysis, the sizes are expressed in units of microlensing Einstein radius $\eta_0$, i.e.,
\begin{equation}
\eta_0 = \sqrt{\frac{4G\avg{M}}{c^2}\frac{D_{\rm {os}}D_{\rm {ls}}}{D_{\rm {ol}}}} \sim 2.03\times10^{16}\,{\rm {cm}}\,\sqrt{\frac{\avg{M}}{0.3\,M_{\sun}}},
\label{equ:RE}
\end{equation}

\noindent where $\avg{M}$~is the average mass of the stars in the lensing galaxy.  
\noindent The different steps of our analysis can be summarized as follows:

\begin{enumerate}[i)]
        
        \item{We use the inverse ray-shooting code of \cite{Wambsganss1998} to generate microlensing magnification pattern of 100$\times$100 $\eta_0^2 \sim $0.657$\times$0.657\,pc$^2$. We fix the value of the local convergence and shear at the location of image $D$ to $(\kappa, \gamma)=(0.576,0.68)$, \citep{Pooley2012}, and calculate maps for ten different fractions $f_*$ of the total matter density in form of compact objects, logarithmically distributed in the range $f_* \in [0.01,1]$. Because of the small dependence of microlensing properties on the mass function of microlenses \citep{Lewis1996, Kochanek2004a}, we assume that all the microlenses have the same mass. We then convolve the maps with the luminosity profile of the source. We model the disc as a two-dimensional Gaussian profile. The exact shape of the profile has little influence on the results \citep{Mortonson2005}, provided that profiles with the same half-light radius $R_{1/2}$ are considered. We consider 51 logarithmically distributed half-light radii $R_{1/2} \in [0.025, 2.5]\,\eta_0$}.
        
        \item {We generate a list of $2\times10^6$ random positions to be extracted from the maps, excluding 500 pixels (5$\eta_0$) at the border of the map. }
        
        \item {Based on those positions, we extract the magnification for each source size, and construct a chromatic curve representing the magnification as a function of the source size.}
        
        \item {The wavelength of observations $\lambda$, and size of the source are related through the equation 
                \begin{equation}
                \sigma = \sigma_0 \left(\frac{\lambda}{\lambda_0}\right)^\nu,
                \label{equ:AD}
                \end{equation}
                \noindent where $\sigma_0$ is the disc size (in practice, the width of a Gaussian profile) at the reference wavelength $\lambda_0$. This relation enables us to transform the dependence of the amplitude of microlensing with the size of the source into a wavelength dependence of the microlensing signal. We compare each simulated microlensing light curve to the data, and calculate a $\chi^2 = \sum{(\mu_{\rm{sim}}-\mu_{\rm{obs}})^2/\sigma^2_{\mu, \rm{obs}}}$, where the sum is over all the observed wavelengths, and $\sigma_{\mu, \rm{obs}}$ is the uncertainty in the estimated microlensing magnification at each wavelength. When necessary, we perform linear interpolation of the simulated microlensing signal to derive the amplitude of microlensing at the wavelength of the observations. } 
        
        \item{We calculate the likelihood of each pair of parameter ($\sigma, \nu$), via summing the likelihood of each light curve estimated as $L_i (F_{\rm{obs}}| \sigma, \nu, f_*)= \exp(-\chi^2/2)$. This enables us to derive, for each fraction of compact matter $f_*$, a two-dimensional map of the likelihood of the source parameters. }
        
        \item {We perform a weighted sum of the likelihoods associated with each $f_*$, using the dark matter probability distribution derived by \cite{Pooley2012}, based on the study of the X-ray microlensing in the system. We then convert the likelihood into probability distribution using the Bayes theorem.  }
        
\end{enumerate}

\subsection{Results}

\begin{figure}
  \includegraphics[width=\hsize]{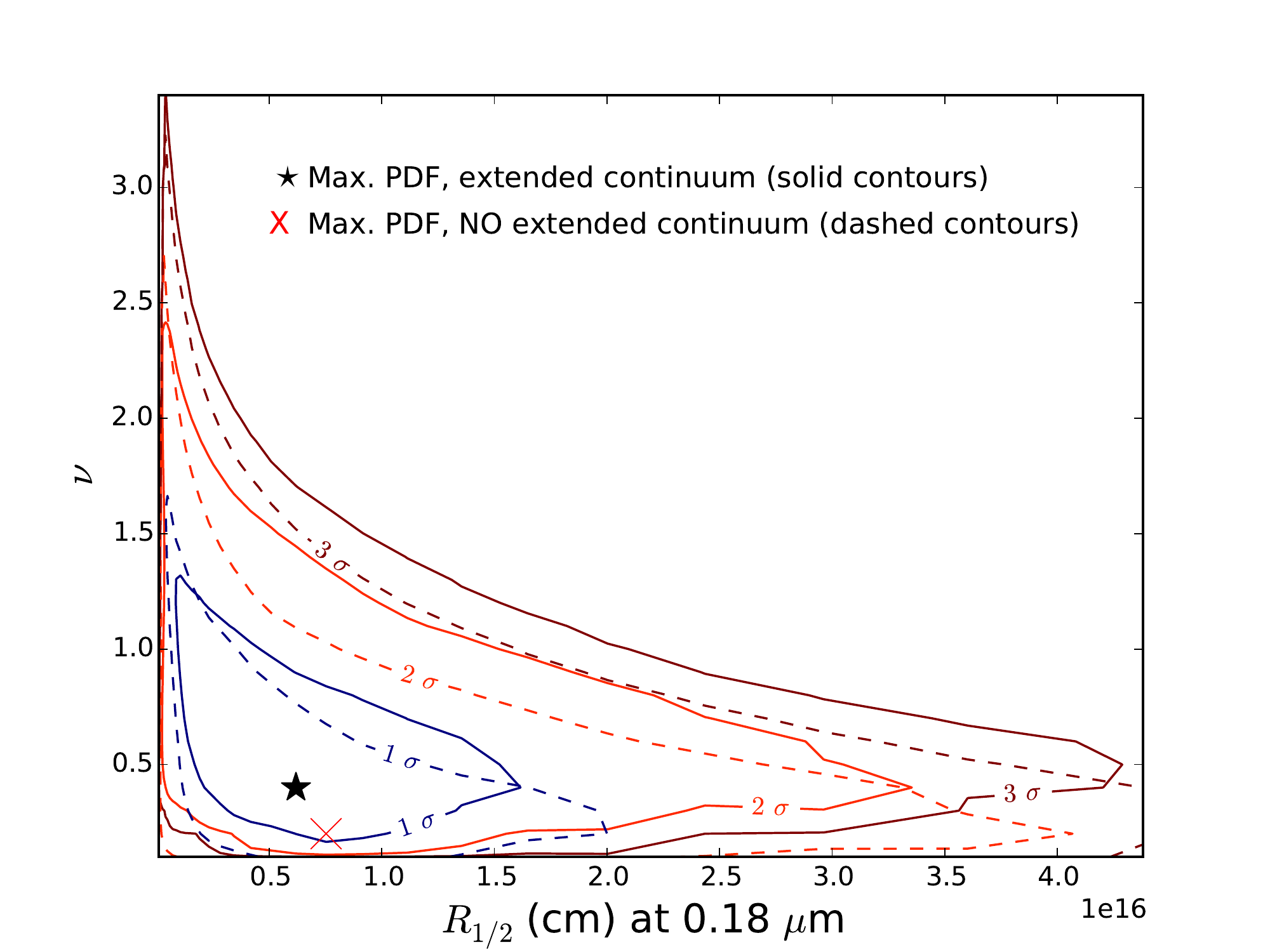}
  \caption{Probability distribution for the size of the disc $R_{1/2}$, and for the power-law index $\nu$ of the temperature profile. Solid contours correspond to the constraints obtained after correction of the microlensing amplitude for the presence of an extended continuum, and the dashed contours show the constraints if we do not account for the extended continuum. Contours are 68.3, 95.5, and 99.7\% confidence level. The corresponding peaks of the probability distribution are shown as a filled black star and red cross, respectively.}  
  \label{fig:PDF}
\end{figure}

\begin{figure}
        \includegraphics[width=\hsize]{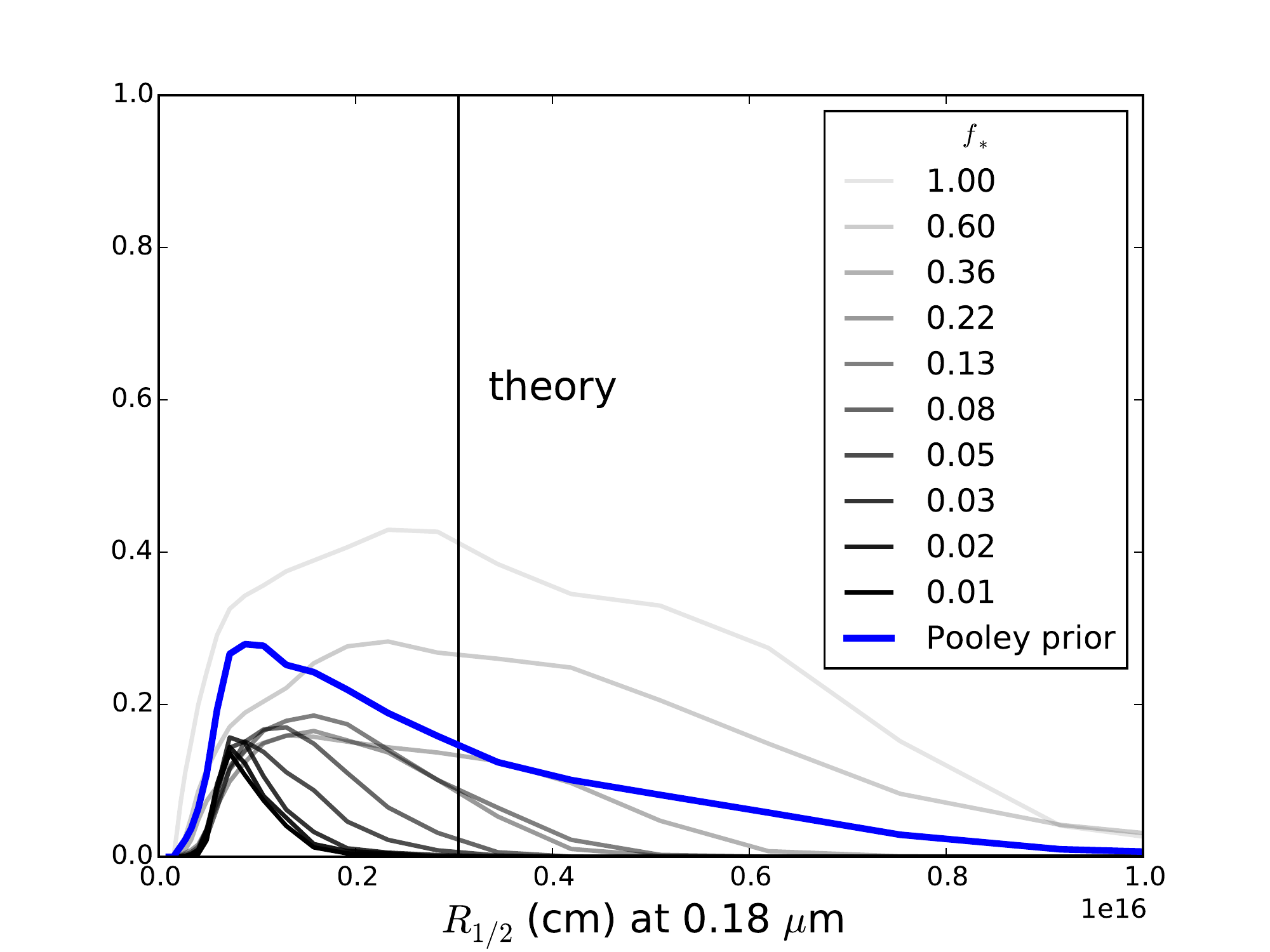}
        \caption{PDF of the source size for a standard accretion disc model ($\nu=4/3$) and various fractions of compact objects $f_*$ (extreme values $f_*$=1, shown in light gray, and $f_*$=0.01, shown in dark gray, correspond to  100\% and 1\% of the total matter being compact, respectively). The thick blue line is the result obtained assuming the same prior on $f_*$, as in Fig.~\ref{fig:PDF}, namely the prior resulting from the analysis of \cite{Pooley2012}. The vertical solid line is the theory size for a Shakura-Sunyaev model based on Eq.~\ref{equ:theory2} for $\cos(i)=0.5$ ($R_{1/2} \sim 3.04\,10^{15}$\,cm $\sim 10^{-3}$\,pc).}
        \label{fig:AD-SS}
\end{figure}

The 2-D probability density function $d^2P/(dR_{1/2}\,d\nu)$ resulting from our analysis is shown in Fig.~\ref{fig:PDF}. This PDF is marginalized over the fraction of compact objects $f_*$ in the lensing galaxy, weighted by the $dP/df_*$ estimated by \cite{Pooley2012}. Our most likely values of the disc size and power-law index of the temperature profiles are $R_{1/2} (\lambda_0 = 0.18\,\mu m)= 0.3\,\eta_0$ = 6.1$\times$10$^{15}\,\sqrt{\avg{M}/0.3\,M_\odot}$\,cm = 0.00197\,$\sqrt{\avg{M}/0.3\,M_\odot}$\, pc, and $\nu = 0.4 $. However, those values significantly depend on our prior on the fraction of compact objects. This is illustrated in Fig.~\ref{fig:AD-SS}, where we show the PDF for $R_{1/2}$ at a fixed value of $\nu = 4/3$ (i.e., the standard accretion disc value), and for various amounts of the compact matter fraction $f_*$. Those PDFs are compared to the case where we use the prior on $f_*$ from Pooley et al. (\citeyear[][i.e., as in Fig.~\ref{fig:PDF}]{Pooley2012}).

Our size estimates can be compared to expectation from theory. The radius $R_\lambda$ at which the disc temperature equals the photon energy, $kT = hc/\lambda_{\rm rest}$ is given by the following expression ~\citep{Poindexter2010b, Blackburne2011a}:
\begin{equation}
R_\lambda = 9.7\times 10^{15} \left(\frac{\lambda}{{\rm{\mu m}}}\right)^{4/3} \left(\frac{M_{BH}}{10^9}\right)^{2/3} \left(\frac{L}{\eta\,L_{edd}}\right)^{1/3} {\rm{cm}} 
\label{equ:theory}
,\end{equation}


\noindent where $\eta$ is the accretion efficiency (assumed to be $\eta=0.1$), $M_{BH}$ the black hole mass, and $L_{edd}$ is the Eddington ratio. An alternative expression for the size which depends only on the magnification corrected luminosity can also be derived (Morgan et al. \citeyear{Morgan2010a}, or based on $T_{{\rm eff}}$ and $\dot{M}$ in Davis and Laor \citeyear{Davis2011} and Laor and Davis \citeyear{Laor2014}):

\begin{equation}
R_\lambda = 3.86\times 10^{15} \left(\frac{L_{\rm {opt, 45}}}{\cos(i)}\right)^{1/2} \left(\frac{\lambda}{{\rm{\mu m}}}\right)^{4/3} {\rm {cm}}
\label{equ:theory2}
,\end{equation}

\noindent where $L_{\rm {opt, 45}}$ is the luminosity at 4861\AA\, in units of $10^{45}$\,erg\,s$^{-1}$\,cm$^{-2}$, and $i$~is the inclination of the disc. Figure~\ref{fig:AD-SS} shows that the theory size \citep[with $L_{\rm {opt, 45}} =$\,5.22;][]{Sluse2012b} is somewhat larger than the microlensing size (using $\avg{M}=0.3\,M_\odot$ in Eq.~\ref{equ:RE}). The presence of an extended continuum  only has a small impact on this result. If we repeat our analysis using microlensing factors derived by forcing $F_M =0$ in the continuum (e.g., $\mu^\prime = 1.75$ in Fig.~\ref{fig:civ}, see also Appendix~\ref{AppendixMmD}), we then find the PDF represented by the dashed contours in Fig.~\ref{fig:PDF}, which peaks at $(R_{1/2}, \nu) = (0.37\,\eta_0, 0.2)$. 

Because we cannot exclude that the amplitude of microlensing in the compact continuum is larger than the value used above (for $F_{ce}/F_{cc} \sim 0.4$), we experimented with two alternative hypotheses for the chromatic behavior of the extended continuum  (Sect.~\ref{subsec:chroma-extended}). On the one hand, we  assumed that in the blue (i.e., up to \CIV), $F_{ce}/F_{cc} \sim 1$, yielding an increase of the amplitude of microlensing in the blue, namely $\mu=2.5\pm0.12$ for \CIV\, and $\mu=2.56\pm0.13$ for $\Lyalpha$. On the other hand, we  considered a constant fraction of extended continuum observed from \CIV\,to \Halpha\, (this yields an increase of the amplitude of microlensing $\mu \sim 2$ around \Hbeta\, and \Halpha). The most likely sizes derived for those alternative signals fell within the 1$\sigma$ solid contours of Fig.~\ref{fig:PDF}. We can then conclude that our estimates of $R_{1/2}$ and $\nu$ are robust within a factor 2 with respect to $F_{ce}/F_{cc}$, and that the strongest source of error is the prior on $f_*$.  

\subsection{Discussion}
\label{MLcontinuum}

In recent years, microlensing studies based on significant samples of microlensed quasars found that the size of the accretion disc is larger by a factor 2 to 10 than the theory size derived from Eq.~\ref{equ:theory} \citep[][and see Hall et al. \citeyear{Hall2014} for a summary]{Morgan2010a,Blackburne2011a, Jimenez2012}. Our snapshot (i.e., single epoch) microlensing measurement of \obj\,instead favors a most likely half-light radius smaller than the theoretical prediction, although the upper limit of the confidence interval extends to about two times the theory size (99.7\% C.I.; Fig.~\ref{fig:AD-SS}). There is therefore no tension between the disc size that we measure in \obj\, and that measured in other systems. More interesting is the impact of the presence of an extended continuum amounting to 30\% of the total flux. Figure~\ref{fig:PDF} shows that the disc size shrinks by typically 20\% when we account for the extended continuum. \cite{Morgan2010a} and \cite{Dai2010a} reached similar conclusions, assuming that 30\% of the broadband flux in lensed quasars  comes from regions of larger physical extent than the compact continuum. 

The other constraint provided by our data regards the power-law index of the disc temperature profile. We find a most likely index, $\nu=0.4$, which is significantly smaller than that of the standard accretion disc, $\nu = 4/3$ (Fig.~\ref{fig:PDF}). Nevertheless, the latter is compatible with our data within 2\,$\sigma$, as generally found for measurements based on single epoch data \citep[e.g.,][]{Bate2008, Floyd2009a}. Our result also agrees with the finding of \cite{Blackburne2011a} that microlensing measurements support a steeper temperature profile than in the standard model. Again, we should underline the rather small impact of the extended continuum on those results. Conversely, the fraction of compact objects in the lens does play an important role on the parameter estimates. As suggested by Fig.~\ref{fig:AD-SS}, our data are more easily reproduced by a standard disc for higher values of $f_*$, although a large fraction of compact objects is not favored by X-ray microlensing studies of \obj~\citep{Pooley2012}. 

In summary, the extended continuum emission is not sufficient  to explain discrepancies generally found between the standard accretion disc model and the microlensing measurements, however, we emphasize the important role played by the prior on the dark matter fraction on size measurements. Dark matter fraction estimates result from analyses, which generally assume the size of the source. \cite{Jimenez2014} has recently demonstrated that this assumption has a non-negligible effect on the derived dark matter fraction at the location of the lensed images. The common assumption of a very small source size (especially in the X-rays) in dark matter fraction inferences yields a fraction of mass in compact form ($f_*$) smaller than what is found by accounting for the extended nature of the source. If we postulate that the prior on $f_*$ from \cite{Pooley2012} is shifted to low values because of this source size effect, we then find that the most likely disc parameters derived from microlensing in \obj\, are closer to those of the standard accretion disc model.

\section{Conclusions}
\label{sec:conclusions}

The technique of quasar microlensing, which consists of analyzing the deformation of gravitationally lensed quasar spectra produced by stars in a lensing galaxy, is a powerful tool for studying the structure of the accretion disc since it constrains the size of the microlensed emission. This method generally shows that the size of the accretion disc is significantly larger than the theory size \citep[see][for a good summary]{Hall2014}. In this work, we have analyzed new spectra (obtained in 2011) of the lensed images $A$ and $D$ of the cloverleaf \obj, a quadruple-imaged broad absorption line system located at redshift $z\sim 2.55$. Because of the presence of absorption covering the source of continuum, and of the existence of microlensing in one of the lensed images, we have unveiled the presence of two sources of continuum: a compact continuum-emitting region that is small enough to be microlensed, and a spatially separated nonmicrolensed continuum-emitting region that accounts for about 30\% of the total continuum in the UV rest-frame part of the spectra (namely 1000-3000\AA). Because this extended continuum is detected in the UV, and is absorbed by the broad absorption trough, we think that it may not come from the host galaxy, but from the innermost regions of the system (possibly as small as the quasar torus or the broad line region). This extended continuum may be scattered light reprocessed either in the same scattering region as that at the origin of the polarization in BAL quasars, or closer to the compact continuum source as in the model proposed by \cite{Lawrence2012}. A reanalysis of archive spectra, shows that this extended continuum was already present in the past 22 years, but was less prominent compared to the compact continuum. 

Owing to our method of analysis of the spectra, we  effectively deblended the compact continuum from the nonmicrolensed continuum. The amplitude of microlensing for the compact continuum, which we attribute to the accretion disc emission,  decreases as the wavelength increases. We use this chromatic variation of the microlensing to derive the size and index of the temperature profile of the disc, assuming that the latter varies like $T \propto R^{-1/\nu}$ (where $\nu = 4/3$ corresponds to the standard Shakura-Sunyaev value). For this purpose, we develop a Bayesian analysis scheme, which compares simulated microlensing signals for different source sizes and index $\nu$ to the observed amplitude of microlensing. This analysis yields a most likely half-light radius of the source of $R_{1/2} = 0.61 \times 10^{16}\,$cm at 0.18\,$\mu$m, and a most likely index $\nu=0.4$. The index is smaller than the prediction from the standard model of accretion, but remains compatible with the latter at the $2\sigma$ level. Fixing $\nu=4/3$, we find a disc size that is about four times smaller than the theory size, but still statistically compatible. There is therefore no evidence for a stronger discrepancy between theory and microlensing disc size in this broad absorption line quasar compared to what is found in other systems. 

To test whether the extended continuum emission could explain the tension between standard theory and microlensing results, we repeated our measurement using an effective amplitude of microlensing, which is the amplitude of microlensing measured including both the compact and the extended continuum. This is equivalent to the measurements performed for other systems, where extended continuum emission is not detected (even when present). As expected, this procedure yields an increase of the microlensing size, and a decrease of the index $\nu$, but the change is not large enough to explain the large microlensing sizes that are commonly reported. We emphasize that another parameter of the analysis, the fraction of matter in compact form in the lensing galaxy, may play a more important role in the source size measurement than the presence of an extended continuum. This is especially true for lensed systems where a small fraction of matter in compact form is assumed in simulations.

Our detection of two spatially separated sources of continuum has been possible owing to differential absorption between the two sources of continuum, and to microlensing of the most compact continuum. The same method may be used in other broad absorption line systems, but its guarantee of success depends on the exact location of the absorbers with respect to the central engine. Innovative methods need to be developed to answer the many open questions about this extended continuum such as: Is it present in other quasars, and visible for all quasar types? How does this continuum emission vary with wavelength? Is the extended continuum scattered light by the BAL wind, by the dust torus, or by cold clouds inner to the BLR?

\begin{acknowledgements}
We are very grateful to Joachim Wambsganss for making available his inverse ray-shooting code used for simulating the microlensing maps of \obj. We thank Ari Laor for enlightening discussions, and the referee for useful comments which helped to improve the presentation of our results. DS acknowledges support from a {\it {Back to Belgium}} grant from the Belgian Federal Science Policy (BELSPO), and partial funding from the Deutsche Forschungsgemeinschaft, reference SL172/1-1. D.H. and L.B. are respectively Senior Research Associate and Research Assistant at the F.R.S.-FNRS. Support for T. Anguita is provided by the Ministry of Economy, Development, and Tourism's Millennium Science Initiative through grant IC120009, awarded to The Millennium Institute of Astrophysics, MAS and proyecto FONDECYT 11130630. This work was supported by the Fonds de la Recherche Scientifique – FNRS under grant 4.4501.05. This research made use of Astropy, a community-developed core Python package for Astronomy \citep{Astropy2013}. 
\end{acknowledgements}

\bibliographystyle{aa}
\bibliography{H1413bib}

\newpage 

\appendix

\section{MmD with an absorbed extended continuum}
\label{AppendixMmD}

We consider a line profile where absorption is present.  We write
\begin{eqnarray} 
F_1 & = & M \, (F_{ee} + F_{ce} \, e^{-\tau_{ce}}) + M \mu \, F_{cc} \, e^{-\tau_{cc}} ,\\ 
F_2 & = & F_{ee} + F_{ce} \, e^{-\tau_{ce}} + F_{cc} \, e^{-\tau_{cc}}
,\end{eqnarray} 
where $F_{ce}$ represents the flux from an extended source of
continuum, which is only macrolensed as the emission line flux
$F_{ee}$,  $F_{cc} \ne 0$ is the flux from the source of continuum
compact enough to be  microlensed, and $\tau_{cc}$ ($\tau_{ce}$) is the optical depth of the absorber in front of $cc$ ($ce$). Using these relations in Eqs. 3
and 4 with $A=M\mu$, we obtain
\begin{eqnarray} 
F_M \ & = &  F_{ee} + F_{ce} \, e^{-\tau_{ce}} + \, \frac{A'-A}{A' - M} \, F_{cc} \, e^{-\tau_{cc}} \\ 
F_{M\mu} & = & F_{cc} \, e^{-\tau_{cc}} + \, \frac{M-M'}{A - M'} \, (F_{ee} + F_{ce} \, e^{-\tau_{ce}}) \; .
\end{eqnarray}
In these equations, $A'$ and $M'$ are parameters tuned to satisfy
constraints on $F_M$ and $F_{M\mu}$. The constraint $F_{M\mu} = 0$ at
wavelengths where $F_{cc} \, e^{-\tau_{cc}} = 0$ and $F_{ee} \ne 0$ can
only be satisfied with $M'=M$.  If $M$ is correctly estimated,
$F_{M\mu} = F_{cc} \, e^{-\tau_{cc}} \; \geq 0$ independently of the
existence of a nonmicrolensed extended continuum.

In the absence of a nonmicrolensed extended continuum (i.e., $F_{ce}
= 0$), making $F_M = 0$ at wavelengths where $F_{ee} = 0$ and
$\tau_{cc}=\tau_{ce} = 0$ (i.e., in the unabsorbed continuum) gives $A'=A$.  In the
presence of a nonmicrolensed extended continuum, making $F_M = 0$ in
the unabsorbed continuum gives $A'\ne A$. In this case, $F_M = F_{ce}
\, (e^{-\tau_{ce}}-e^{-\tau_{cc}})$ at wavelengths where the continuum
is absorbed and $F_{ee} = 0$, which can result in either positive or
negative values for $F_M$ whether $\tau_{ce} < \tau_{cc}$ or $\tau_{ce} >
\tau_{cc}$, respectively. Making $F_M \geq 0$ allows the detection of the
extended continuum in the latter case ($\tau_{ce} > \tau_{cc}$), but $A'= A$
is derived only if $e^{-\tau_{ce}}=0$ at some wavelengths where $F_{ee}
= 0$ and $e^{-\tau_{cc}} \ne 0$. Without absorption (i.e., $\tau_{ce} =
\tau_{cc} = 0$), such a nonmicrolensed extended continuum would remain
unnoticed.

Note that $\mu^\prime = A^\prime/M$,  determined by making $F_M = 0$ in the
unabsorbed continuum, can be seen as the average microamplification of
the whole continuum source, while $\mu = A/M$ refers to the
microamplification of the compact continuum source. They are related
by
\begin{eqnarray} 
\frac{\mu  -1}{\mu^\prime -1} &  = &  1 + \frac {F_{ce}}{F_{cc}}  \; .
\label{eq:fracCC}
\end{eqnarray}

\section{Equivalent width in the absorption trough}
\label{AppendixEW}

In order to understand how microlensing affects the EW, we can write Eq.~\ref{equ:EW}, with more explicit expression of $F_\lambda$ and of the continuum flux $\mathcal{F}$. We assume that the latter is constant over the emission line and the absorption trough and is well approximated by the continuum flux measured in the vicinity of the line. We consider a broad emission line characterized by a flux $F_{\lambda,E}$, and a continuum originating from two regions, a compact region $cc$, compact enough to be microlensed, and a more extended region $ce$, which is not microlensed. We can specialize Eq.~\ref{equ:EW} in the presence of absorption. For a lensed image $i$, we obtain 

\begin{equation}
EW_i = \int_{\lambda_1}^{\lambda_2} \left(1-\frac{\mu_i\,e^{-\tau_{cc}}\,F_{\lambda,cc}+e^{-\tau_{ce}}\,F_{\lambda,ce}+e^{-\tau_{E}}\,F_{\lambda,E}}{\mu_i\,F_{cc}+F_{ce}}\right) \,d\lambda,
\label{equ:EWi}
\end{equation}

\noindent where $\mu_i$ is the amplitude of micromagnification of the compact continuum $cc$, and $\tau_{cc}$, $\tau_{ce}$, and $\tau_E$ are the optical depths of the gas absorbing the compact continuum, the extended continuum, and the broad emission line, respectively. Contrary to Appendix~\ref{AppendixMmD}, we make explicit the fact that the emission line can be absorbed (i.e., $F_{ee} = e^{-\tau_{E}}\,F_{\lambda,E}$). 

Equation~\ref{equ:EWi} clearly shows that if $F_{\lambda, ce} = F_{\lambda, E} = 0$ (i.e., no extended continuum or emission line), and if the continuum does not vary over the range $[\lambda_1, \lambda_2]$ (i.e., $F_{\lambda, cc} \approx F_{cc}$), then the EW is independent of the amplitude of microlensing $\mu_i$. In the following, we compare the difference of equivalent widths $\Delta EW = EW_D-EW_A$, between image $D$, characterized by microlensing $\mu$, and image $A$, unaffected by microlensing, in the presence of flux from the emission line and/or from an extended continuum region. 

\subsection{Compact continuum+emission line}

We first consider the situation where there is only the compact source of continuum $F_{cc}$ and the emission line over the wavelength range $[\lambda_1, \lambda_2]$. We then have $F_{\lambda, ce} = 0$ in ~\ref{equ:EWi}. Assuming that the continuum flux does not vary over the range $[\lambda_1, \lambda_2]$ such that $F_{\lambda,cc} = F_{cc}$, we can rewrite ~\ref{equ:EWi}: 

\begin{eqnarray} 
EW_i &=\frac{1}{\mu_i\,F_{cc}} \int_{\lambda_1}^{\lambda_2}
        \left(\mu_i\,F_{cc}-\mu_i\,e^{-\tau_{cc}}\,F_{cc}-e^{-\tau_{E}}\,F_{\lambda, E}\right)\,d\lambda \\
        & =\int_{\lambda_1}^{\lambda_2} \left(1-e^{-\tau_{cc}}\right)\,d\lambda - \frac{1}{\mu_i\,F_{cc}}\,\int_{\lambda_1}^{\lambda_2} \left(e^{-\tau_{E}}\,F_{\lambda, E}\right)\,d\lambda
.\end{eqnarray}


The difference of equivalent width, $\Delta EW = EW_D-EW_A$, can be derived from the above equation specialized for the pair of images, i.e.,
\begin{eqnarray} 
 \Delta EW = \frac{\mu-1}{\mu}\, \frac{1}{F_{cc}}\,\int_{\lambda_1}^{\lambda_2} \left(e^{-\tau_{E}}\,F_{\lambda, E}\right)\,d\lambda.
\label{equ:DEW_FE}
\end{eqnarray}

It results from this equation that when $\mu > 1$, one may only observe $EW_D > EW_A$.
 
\subsection{Compact + extended continua}

We now consider that there is no flux from the emission line ($F_{\lambda,E} = 0$) over the wavelength range $[\lambda_1, \lambda_2]$, but two sources of continuum: a compact continuum microlensed by a factor $\mu_i$ and an extended continuum $F_{ce}$ which is not microlensed. Assuming that the continuum flux is constant over the range $[\lambda_1, \lambda_2]$ such that $F_{\lambda,cc} = F_{cc}$ and  $F_{\lambda,ce} = F_{ce}$, we obtain
\begin{eqnarray} 
  EW_i &=\frac{1}{\mathcal{F}_i} \,\int_{\lambda_1}^{\lambda_2}
  \left(\mu_i\,F_{cc}+F_{ce}-\mu_i\,e^{-\tau_{cc}}\,F_{cc}-e^{-\tau_{ce}}\,F_{ce}\right)\,d\lambda \\
   &=\frac{1}{\mathcal{F}_i} \int_{\lambda_1}^{\lambda_2} \left[\mu_i\,F_{cc}\left(1-e^{-\tau_{cc}}\right) +  \left(e^{-\tau_{ce}}\,F_{ce}\right)\right]\,d\lambda
,\end{eqnarray}
\noindent where we  defined $\mathcal{F}_i = \mu_i\,F_{cc}+F_{ce}$, the flux in the unabsorbed continuum. 

The difference of equivalent width $\Delta EW = EW_D-EW_A$, is then

\begin{eqnarray} 
\Delta EW = \frac{F_{cc}\,F_{ce}\,(\mu-1)}{\mathcal{F}_D \, \mathcal{F}_A}\,\int_{\lambda_1}^{\lambda_2}  \left(e^{-\tau_{ce}}-e^{-\tau_{cc}}\right)\,d\lambda.
\label{eq:DEW_Fce}
\end{eqnarray}

We clearly see that a difference of equivalent width is only  detected if there is a difference of opacity between the absorbers covering the extended and the compact continua, i.e., $\tau_{cc} \neq \tau_{ce}$. If $\mu > 1$, we find $\Delta EW > 0 $ ($EW_D > EW_A$), and if $\tau_{cc} > \tau_{ce}$, and conversely $\Delta EW < 0 $ ($EW_A > EW_D$) if $\tau_{cc} < \tau_{ce}$. 

\subsection{Compact + extended continua + emission line}

The situation where the source of light under the absorption is the sum of a compact and an extended continuum and of an emission line, directly results from the two previous cases. If we define the flux of the unabsorbed continuum of image $A$ and $D$, $\mathcal{F}_A = F_{cc}+F_{ce}$, and $\mathcal{F}_D = \mu\,F_{cc}+F_{ce}$, then we can write the following difference of equivalent width $\Delta EW = EW_D-EW_A$ between images $D$ and $A:$
\begin{eqnarray}
\Delta EW = \frac{F_{cc}\,(\mu-1)}{\mathcal{F}_D \, \mathcal{F}_A}\,\int_{\lambda_1}^{\lambda_2} \left[F_{ce}\,\left(e^{-\tau_{ce}}- e^{-\tau_{cc}}\right) + e^{-\tau_{E}}\,F_{\lambda, E}\right]\,d\lambda.
\label{equ:EWall}
\end{eqnarray}

From this equation, we see that when $\mu >1$ and $\tau_{cc} > \tau_{ce}$ (i.e., when the compact continuum is more absorbed than the extended continuum), $\Delta EW > 0$. Conversely, $\Delta EW < 0$ can only be obtained if  $\tau_{cc} < \tau_{ce}$, i.e., when the compact continuum is less absorbed than the extended continuum. Positive value of  $\Delta EW$ may also occur depending on the absorption rate of the different components, and on the relative contribution of the extended continuum and of the emission line.

\end{document}